\documentclass[12pt,preprint]{aastex}

\shorttitle{Loop widths and scale heights}
\shortauthors{Petrie}

\begin{document}

%% LaTeX will automatically break titles if they run longer than
%% one line. However, you may use \\ to force a line break if
%% you desire.

\title{Coronal loop widths and pressure scale heights}

%% Use \author, \affil, and the \and command to format
%% author and affiliation information.
%% Note that \email has replaced the old \authoremail command
%% from AASTeX v4.0. You can use \email to mark an email address
%% anywhere in the paper, not just in the front matter.
%% As in the title, use \\ to force line breaks.

\author{G.J.D. Petrie}
\affil{National Solar Observatory, Tucson, AZ 85719}

%% Notice that each of these authors has alternate affiliations, which
%% are identified by the \altaffilmark after each name.  Specify alternate
%% affiliation information with \altaffiltext, with one command per each
%% affiliation.

%\altaffiltext{1}{} 
%\altaffiltext{2}{}

%%%%%%%%%%%%%%%%%%%%%%%%%%%%%%%%%%%%%%%%%%%%%%%%%%%%%%%%%%%%%%%%%%%%%%%%%%%%%%%%%
\begin{abstract}
The scale heights of stratification and the widths of steady solar coronal loops exhibit properties unexplained by standard theory: observed scale heights are often much greater than static theory predicts, while the nearly-constant widths of loop emission signatures defy theoretical expectations for large flux tubes in stratified media.  In this work we relate the cross-sectional profile of a coronal flux tube to its density scale height in steady-state plasma flow regimes.  Steady flows may shorten or lengthen the scale height according to how the tube cross-sectional area varies with arclength.   In a near-potential corona the flux tubes are expected to be sufficiently expansive in many active regions for scale heights to be increased by steady flows.  On the other hand, cases where scale lengths are actually increased to observed sizes form a small part of the solution space, close to regimes where density profiles reverse.  Therefore, although steady flows are the only steady process known to be capable of extending scale heights significantly, they are not expected to be not responsible for the majority of extended active region scale heights.
\end{abstract}


%%%%%%%%%%%%%%%%%%%%%%%%%%%%%%%%%%%%%%%%%%%%%%%%%%%%%%%%%%%%%%%%%%%%%%%%%%%%%
\section{Introduction}

In the solar corona, closed plasma loops are believed to trace lines of magnetic force that penetrate the photosphere from below and expand to fill the coronal volume above an active region (Bray et al.~1991, Aschwanden~2004).  They are curved filamentary structures connecting regions of opposite magnetic polarity.  Plasma loops are often observed as isolated brightened flux tubes embedded in a less emissive but more homogeneous field configuration.  A broad consensus has developed that coronal loops are nearly isothermal along their arclengths, while temperatures and densities can vary on small scales between neighboring loop structures (Rosner et al.~1978, Lenz et al.~1999, Aschwanden et al.~1999, 2000, Aschwanden and Nitta~2000, Schmelz et al.~2001, Martens et al.~2002, Schmelz~2002, Winebarger et al.~2002).  This can be explained by the thermal conductivity being much more efficient along than across the magnetic field in the nearly fully-ionized coronal gas (Spitzer~1962).

Two physical parameters which continue to cause controversy are the scale height of hydrostatic stratification and the width of the observed loop-shaped emission patterns.  Loop emission in active regions is frequently clearly visible at heights several times greater than the hydrostatic scale height associated with the temperature of emission (Aschwanden et al.~2001).  Furthermore, the widths of these emission signatures remain remarkably constant along their entire lengths.  These emission patterns are usually identified with magnetic flux tubes because of the efficient field-aligned thermal conductivity of the corona, and they therefore defy theoretical expectations that flux tubes must expand in a stratified magnetic field (Lopez Fuentes et al.~2006).  In this work we ask how steady flows can affect scale heights in the expanding flux tubes of the stratified solar corona.

The scale heights of steady loops have already been studied using steady magnetohydrodynamic (MHD) theory by Petrie et al.~(2003) and Gontikakis et al.~(2005) (henceforth P03 and G05) and via a steady hydrodynamical (HD) simulation by Patsourakos et al.~(2004) (henceforth P04).  Intriguingly, these two approaches gave opposite conclusions about the influence of steady flows on scale heights.  While P04 concluded that plasma flows reduce the pressure scale height relative to the static case, P03 and G05 found that the opposite can be true: that steady flows can push the pressure scale height above that of the static case to typical observed values.  In this work we reconcile these results and address the question of what effect steady flows may have in the corona.

\section{Flux tube expansion and conservation of magnetic and mass flux}

The variation of cross-sectional area along a loop is particularly important to understanding plasma dynamics along loops.  P03, P04 and G05 recently conducted different studies of how steady plasma flows along loops may influence the observed fall-off of plasma density, and therefore emission intensity, with increasing height.  P04 assume that loop cross-sectional areas are constant or nearly constant along loops, invoking observations by Klimchuk~(2000) and Watko \& Klimchuk~(2000).  Traditionally HD models have assumed a constant cross-section (Orlando et al.~1995a, b; Reale \& Peres 2000, Winebarger et al.~2003).  Cargill \& Priest~(1980, 1982) investigated the effects of varying flux tube cross-sections.

Such a constant cross-sectional profile is not to be expected of a flux tube in a stratified magnetic field, where the solenoidal condition on the magnetic field $\bf B$, ${\bf\nabla\cdot B}=0$, requires that

\begin{equation}
\frac{d}{ds} (AB)=0,\ AB=\Phi ={\rm constant},
\end{equation}

\noindent where $B(s)$ and $A(s)$ are the magnetic field strength and flux tube cross-sectional area respectively as functions of arclength and $\Phi$ is the total magnetic flux along the tube.  From this relation it is clear that if a tube extends to a height of order the magnetic scale height, the field strength cannot be approximated by a constant and so neither can the tube cross-sectional area.  While it can be argued that within active regions current systems may take the field far from potential states and inhibit the fall-off of the field strength with height, this is surely not the case for the flux tubes corresponding to loops of height of order 100~Mm seen to persist in active regions (e.g. Aschwanden et al.~2000).  It would take enormous currents to influence such tubes significantly (recall that the linear force-free parameter $\alpha$, where ${\bf\nabla\times B}=\alpha {\bf B}$, has dimension $1/L$ where $L$ is the length scale).  While this scenario may occur sometimes, it seems implausible that all or most tall tubes are thus influenced by currents.

In steady HD and MHD the mass conservation equation,

\begin{equation}
\frac{d}{ds} ({\rho vA}) = 0,\ \rho vA={\rm constant},
\label{masscons}
\end{equation}

\noindent where $\rho$ is the plasma density and $v$ the field-aligned velocity, can be integrated to give the result that the mass flux density $\rho v$ is also inversely proportional to the cross-sectional area $A$ along the flux tube, i.e. the magnetic flux and the mass flux are directly proportional along the tube.  Therefore, if the gas density falls off with height, $vA$ must increase with height.  If $A$ is (nearly) constant, this implies a velocity magnitude which increases with height as in P04.  On the other hand, P03 and G05 presented examples with velocity magnitudes decreasing with height.  For this to happen, the magnetic field strength must fall off with height more quickly than the density (independently of the cross-section profile).  This is a much stronger condition than the magnetic pressure falling off more quickly than the gas pressure, which applies in the upper corona.  If the field strength does not decrease sufficiently quickly, the velocity will increase with height (see the unscaled examples in Petrie et al.~2002).

\section{Momentum conservation and scale heights}

The conservation of momentum along a flux tube is described by

\begin{equation}
\rho v\frac{dv}{ds} = -\frac{dp}{ds} - \rho g_{||} ,
\label{momentum}
\end{equation}

\noindent where $g_{||}=g dz/ds$ is the component of the gravitational acceleration parallel to the tube, $p(s)$ is the gas pressure, $z(s)$ is the height of the tube above the base of the model as a function of arclength, $g=g_sR^2/[R+z(s)]^2$, $R$ is the solar radius and $g_s$ is the gravitational acceleration on the solar surface.  Near a point $s=s_0$ on the tube the plasma temperature and gravitational acceleration may be assumed constant.  For the static case $v\equiv 0$, the gas pressure may be approximated near $s=s_0$ by

\begin{equation}
p(s)=p_1\exp \left( \frac{-z(s_0)}{H_0}\right) ,
\end{equation}

\noindent where $H_0=p(s_0)/\rho (s_0)g_0$ is the local hydrostatic scale height, $p_1$ is a constant and $g_0=g(s_0)\approx 274$~m/s$^2$ is the solar surface gravitational acceleration.  Since efficient field-aligned thermal conductivity ensures that flux tubes are nearly isothermal (see Introduction), and the gravitational acceleration does not vary much along tubes of height significantly smaller than $R$, this scale height is characteristic of the tube as a whole and is therefore expected to be reflected in loop emission patterns.  However, this is often not the case.  Steady active regions often exhibit scale heights much larger than this hydrostatic scale height: see Aschwanden et al.~(2001), in particular their Figure 8.  The measured scale heights of active region loops are often 2-4 times larger than the hydrostatic scale heights associated with their temperatures (Aschwanden et al.~2001, Aschwanden~2004).

P03, P04 and G05 have shown that steady plasma flows can alter this scale height significantly.  However, the conclusions of these two studies differed markedly:  P03 and G05 concluded that super-hydrostatic scale heights were possible in steady MHD models while P04 found that steady HD flows were unable to explain the observed superhydrostatic scale heights.  The crucial difference between the two models is that velocity magnitudes decrease with height in the P03 and G05 models while they increase in the P04 model.  The significance of this difference may be understood by examining Equation~\ref{momentum}.

In the left leg of each model, the pressure gradient force $-dp/ds>0$ and the weight $\rho g_{||}<0$.  In the static case these forces cancel.  In P04, $\rho v dv/ds>0$ so that $|dp/ds|>|\rho g_{||}|$, while the opposite is true in P03 and G05.  In the right leg, $-dp/ds<0$ and $\rho g_{||}>0$, and P04, $\rho v dv/ds<0$ so that again $|dp/ds|>|\rho g_{||}|$ and the opposite is again true in P03 and G05.  Velocities increasing with height as in P04 introduce an upward field-aligned inertial force component along each leg of the loop, thereby adding a further force for the gas pressure gradient to balance besides the plasma weight.  In this case an enhanced pressure gradient results and so the pressure scale height becomes smaller than that of the static case.  On the other hand, the expanding flux tubes of P03 and G05 cause the velocity magnitudes to decrease with height, giving a downward inertial force component along each leg, a smaller pressure gradient and a longer scale height.

Whether such a velocity pattern might be typical of coronal flux tubes or not and whether these velocities can routinely influence active-region scale heights in reality are questions taken up in the following sections.  The above considerations demonstrate that the scenario is at least plausible.

\section{Flux tube widths in quiet and active coronal fields}
\label{AltschulerNewkirk}

A clue to whether coronal scale heights might be lengthened in reality is gained by estimating the rates of fall-off with radius of the simplest possible quiet-sun and active-region coronal fields.  Under the simplifying assumption that there are no significant electric currents in the corona, the coronal magnetic field may be represented by ${\bf B}=-{\bf\nabla}\psi(r,\theta ,\phi )$ in spherical coordinates where (Altschuler \& Newkirk~1969)

\begin{eqnarray}
\psi & = & R\sum_{n=1}^{\infty}\sum_{m=0}^{n}\left[ (R/r)^{n+1}\right.\times\nonumber\\
& & \times\left.\left( g_n^m\cos (m\phi )+h^m_n\sin (m\phi )\right) P_n^m(\theta )\right] .\label{expansion}
\end{eqnarray}

\noindent Here the associated Legendre polynomial $P_n^m(\theta )$ is an $n$th-order polynomial of $\sin (\theta)$ and $\cos (\theta )$ and $r=R$ is the solar surface.  (The ``source surface'' introduced by Altschuler \& Newkirk is not relevant to our argument and has been omitted here for simplicity.)  The $g_n^m$ and $h_n^m$ coefficients refer to the multipole components of the magnetic field, e.g. the $g_1^0$ term is the standard polar dipole and the $g_1^1$ and $h_1^1$ terms are the two orthogonal equatorial dipole components.  Higher-order terms refer to the quadrupole, octupole and higher moments.  The principal index, $n$, is the total number of circles of nodes on the sphere for that multipole while the second index $m$ is the number of those nodal circles passing through the pole (Hoeksema~1984).  Therefore small, intense features in the corona must be represented by high-order multipole terms.

For example, a simple quiet-sun polar dipole field  varies on the scale of the solar meridian half-circumference, about $2\times 10^6$~km, and is dominated at these scales by the $n=1$ terms of Equation~(\ref{expansion}) which clearly decrease with radius as $1/r^3$.  The low-order terms such as this dipole dominate the magnetic field in quiet regions of the model (Hoeksema~1984).  On the other hand, the local magnetic field of an active region of characteristic length scale 100~Mm, about 20 times smaller is dominated by terms of order $n=20$ and higher.  Such an active region field will be weaker at height $r=R+50$~Mm than at $r=R$ by a factor of about $1.072^{22}\approx 4.6$, while a hydrostatic plasma at about 1~MK with scale height 50~M will fall off over the same range by a factor of $e\approx 2.7$.  Meanwhile the dipole field falls off by only $1.072^3\approx 1.2$.  This implies that active regions typically have sufficiently expansive flux tubes for steady flows to increase scale heights, while quiet regions do not.

Of course many active region field structures are far from being well described by potential models.  What the potential model does capture is the sharp fall-off of intense active region field strengths as the flux tubes expand to fill the coronal volume above.  While local current systems may influence the structure of active regions, potential models adequately describe flux tube expansions associated with active regions more often than not (Schrijver et al.~2003).

\section{Isothermal steady flow in a potential magnetic loop}

To determine whether steady flows may routinely extend scale heights to observed values under the conditions described above, we give a simple analytical model of steady, isothermal plasma flow through a potential loop-shaped magnetic flux tube.  We model a loop-shaped magnetic flux tube using a well-known potential field solution.  We represent $\bf B$ by using a magnetic flux function
(per unit length in the ${\bf\hat{y}}$ direction) as well as a potential function (also per unit length) in dimensionless coordinates $(x,y,z)$

\begin{equation}
{\bf B}= {\bf\nabla} \alpha (x,z)\times{\bf\hat{y}}=\nabla\psi (x,z) .
\end{equation}

\noindent The field

\begin{equation}
{\bf B} = B_0\left[ C_1\cos (x), 0, -C_1\sin (x)\right] \exp (-z)\label{B2dlff},
\end{equation}

\noindent is a potential (current-free) field with potential $\psi =C_1\sin (x)\exp (-z)$ and flux function $\alpha (x,z)=C_1\cos (x)\exp (-z)$.  Field lines coincide with level curves of $\alpha (x,y)$.  The equation
for the field line defined by $\alpha =\alpha_0$, for $\alpha_0$ constant, is found to be

\begin{equation}
z=\log [C_1\cos (x)]-\log \alpha_0 .\label{fieldlineeq}
\end{equation}

\noindent It can be seen from Equation
(\ref{fieldlineeq}) that two field lines defined by $\alpha =\alpha_1$
and $\alpha =\alpha_2$, for $\alpha_1$ and $\alpha_2$ constants, differ from each other only by a vertical
translation, and that for any point $(x,z_1)$ on the first field line,
the corresponding point on the second field line $(x,z_2)$ can be found
from it by moving vertically a distance,

\begin{equation}
z_2 -z_1 =\log{\frac{\alpha_1}{\alpha_2}} .
\label{verticaltrans}
\end{equation}

\begin{figure}
\begin{center}
\resizebox{0.79\hsize}{!}{\includegraphics*{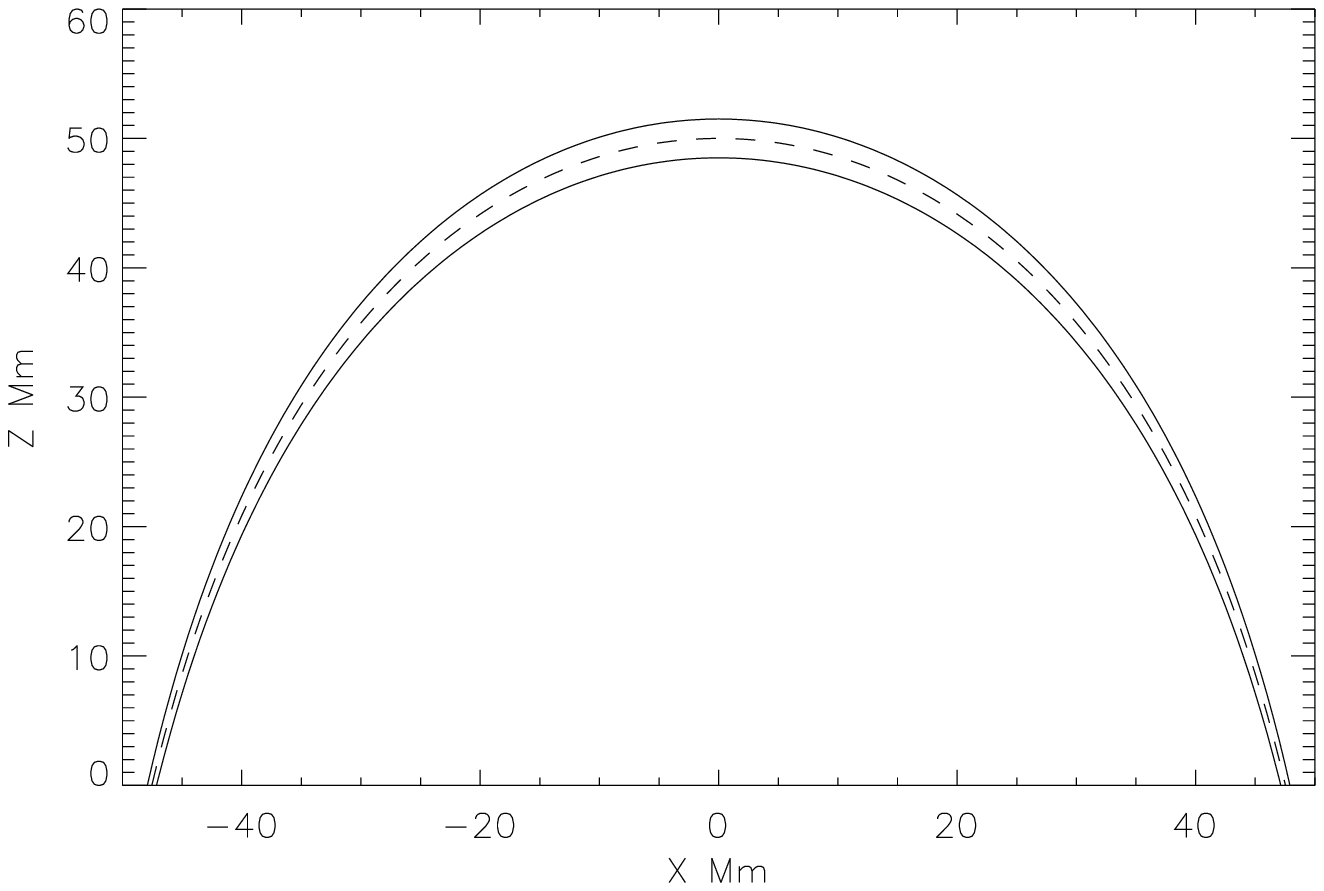}}
\end{center}
\caption{Flux tube formed by two field lines (solid lines) of the spatially self-similar potential field described in the text.  The flux tube central axis is represented by the dashed line.  In this example the flux tube width increases exponentially by a factor of 4 between the foot points at $Z=0$~Mm and the apex at $Z=50$~Mm.}
\label{tube}
\end{figure}

\noindent This property is called spatial self-similarity.  Flux tubes are therefore formed by two field lines that differ by a fixed vertical distance along their entire lengths.  Such a tube is plotted in Figure~\ref{tube}.

The scaled coordinates are given in terms of the dimensionless coordinates by $(X,Y,Z)=Z_0(x,y,z)$, where $Z_0$ is a constant length scale describing the expansiveness with height of the flux tube.  If the flux tube expands by a factor of $n$ over a distance of $D$~Mm, the length scale $Z_0$ is constrained to be $Z_0=D/\log n$~Mm.  This length scale is not to be confused with the hydrostatic scale height $H$.  As discussed in Section~\ref{AltschulerNewkirk} and in Klimchuk~(2000) and Lopez Fuentes et al.~(2006), a reasonable area expansion factor is $n\approx 4$.  We take the tube height to be 50~Mm.  This is the case shown in Figure~\ref{tube} and used in the hydrodynamical model that follows.

\begin{figure}
\begin{center}
\resizebox{0.49\hsize}{!}{\includegraphics*{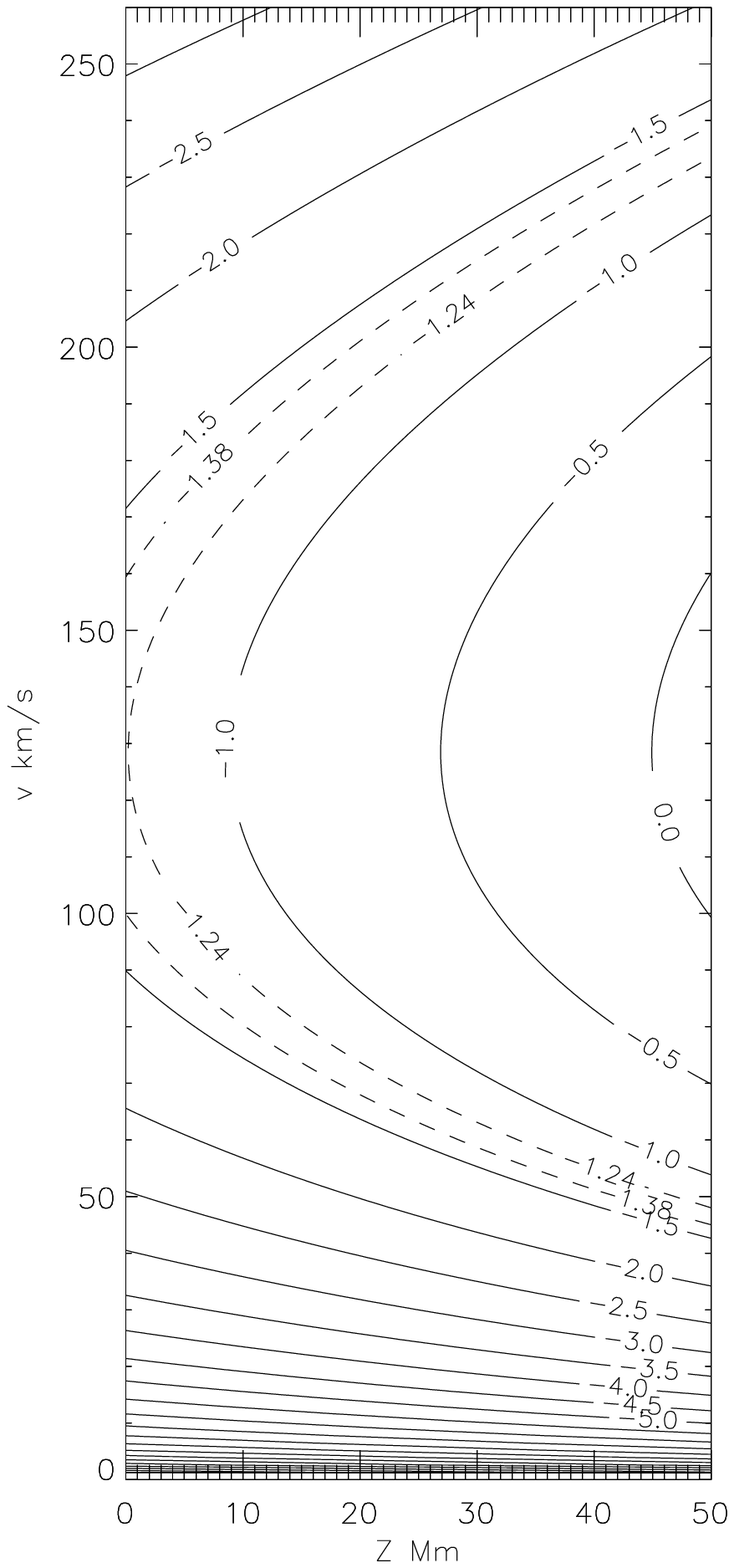}}
\resizebox{0.49\hsize}{!}{\includegraphics*{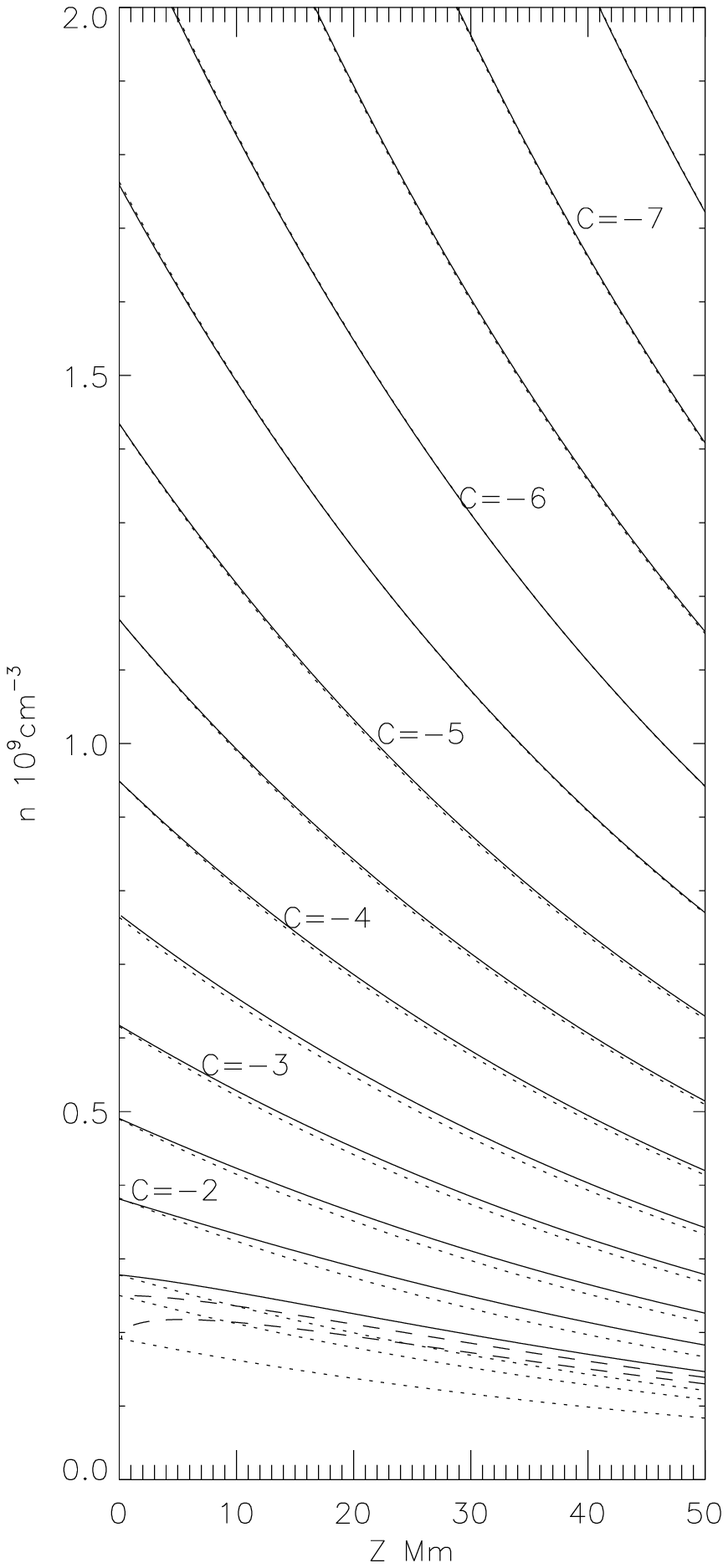}}
\end{center}
\caption{Solution curves for the velocity (left picture) and density (right picture) of the steady hydrodynamical model.  The two critical solutions associated with the isothermal sound speed and the density reversal are represented by dashed lines: contours -1.24 and -1.38 respectively in the left picture and the lower and upper dashed lines respectively in the right picture.  In the right picture, static density profiles are represented by dotted lines.  For each solution, the density magnitude can be scaled freely: the solutions plotted equal mass flux $\rho v A$ for ease of presentation.  Cases where $dv/dz$ become infinite (contours of value $\ge -1.24$) are unphysical.}
\label{vandrho}
\end{figure}

An isothermal steady plasma flow through this flux tube may be modeled by a solution of Equation~(\ref{momentum}) with the flux tube expansion taken into account.  Thus $\rho (s)$, $v(s)$ and $A(s)$ must satisfy Equation~(\ref{masscons}) with $A(s)=A_0 \exp (-Z(s)/Z_0)$ for a constant $A_0$.  We study here cases where loop cross-sectional areas vary exponentially with height, slightly different from Cargill \& Priest's (1980, 1982) cases with nonuniform width where areas vary linearly with arclength.  We are motivated to do this by the form of the potential flux tube and by our interest in variations of loop cross-sectional area with height in this work.  Thus Equation~(\ref{momentum}) becomes

\begin{equation}
\left( v-\frac{c_s^2}{v}\right) \frac{dv}{ds} = \left( \frac{c_s^2}{Z_0} -g_s\right)\frac{dZ}{ds} 
\label{Parkermb}
\end{equation}

\noindent where $c_s=\sqrt{k_BT_0/m}$ is the isothermal sound speed, $k_B$ is Boltzmann's constant and $m$ is the mean particle mass, taken to be half a proton mass in a pure, fully ionized hydrogen plasma.  In this equation, the temperature and the flux tube expansion factor are represented by $c_s$ and $Z_0$ respectively.  For $T_0=10^6$~K as in the model presented here, $c_s\approx 129$~km/s.  For a $10^6$~K plasma in a tube with this expansion factor, $c_s^2/Z_0>g_s$ so that the velocity of any subsonic flow decreases with height, as anticipated in Section~\ref{AltschulerNewkirk}, and scale heights are therefore lengthened.  Examples with less expansive tubes and/or cooler plasma such that $c_s^2/Z_0>g_s$ would have a shortened scale length.  A first integral can be found on eliminating $s$:

\begin{equation}
Z(v)=\frac{v^2/2-c_s^2\log (|v|/c_s)}{c_s^2/Z_0-g_s} +CZ_0,
\end{equation}

\noindent for $C$ a constant.  Solution curves are shown in Figure~\ref{vandrho} (left picture) along with the corresponding solutions for the density (right picture).  These solutions are independent of the specific shape of the flux tube, depending only on the expansion of the tube with height.  The solutions clearly have singular points at $v=c_s$, where $dv/dz$ becomes infinite.  No regular transonic solutions exist in this class, while solutions with shocks may be possible as in Cargill \& Priest~(1980, 1982), whose study also included regular subsonic cases and smooth transonic flows.  (Note that for less expansive flux tubes where velocities increase with height up the left leg, a singular regular transonic solution can exist with sonic point at the apex and a supersonic downflow in the right leg, for in this case the root in the left-hand side of Equation~(\ref{Parkermb}) can be balanced by $dz/ds=0$ at the apex.  More general MHD solution classes also contain regular transonic solutions: see Goedbloed \& Lifschitz~1997, Vlahakis \& Tsinganos~1998 and Petrie et al.~2002.)   Regular subsonic solutions are the most relevant subset for the study of steady loops and we focus on that subset here.

Note that non-zero $z$-derivatives at the apex do not imply discontinuities or kinks in the physical parameters at the apex: variations with arclength are given by $d/ds=d/dz\times dz/ds$ where $dz/ds$ is obviously zero at the apex.  The solutions for $v$ are symmetric in the $z$-axis since the force balance is insensitive to the direction of fluid flow (e.g. upward accelerations and corresponding downward decelerations give the same inertial force).  The inclusion of an energy equation would remove this ambiguity but we avoid that complication here.

% noting only that Equation~(11) of Petrie et al.~(2003) with $\gamma=1$ describes the energy flux associated with this isothermal flow.

The density curves are plotted along with their static equivalents $\rho =\rho_0\exp (-Z/H)$ where $H=k_B T_0/(mg_s)$ is the gravitational scale height.  This is about 60~Mm where we assume a fully ionized hydrogen plasma as in the hydrodynamical model.  (References such as Aschwanden et al.~(2001) and Aschwanden~(2004) quote a value of about 47~Mm assuming a 10:1 ratio of hydrogen to helium.  We model a pure hydrogen plasma here for simplicity.)

The density gradient is given by

\begin{equation}
\frac{d\rho}{dz} =\frac{1}{A_0\exp (z/Z_0)}\frac{v^2/Z_0-g_s}{v^2-c_s^2} .
\end{equation}

At the critical point $v=c_s$ this gradient becomes infinite.  A further critical point arises when $v=\sqrt{g_sZ_0}$, above which speed the density gradient becomes positive and we have a density inversion - unlikely to be a steady phenomenon in the solar atmosphere.  Since $c_s$ and $\sqrt{g_sZ_0}$ are dependent on completely different physical parameters, the former on the temperature and the latter on the flux tube expansion, these two critical velocities are independent of each other.  The sonic point may occur at a lower velocity than the density inversion if the plasma is cooler or the flux tube less expansive than in the example presented here.

Aschwanden~(2004) summarizes the measurements of coronal loop flow velocities using SoHO/CDS, SoHO/SUMER and TRACE by a variety of methods.  Apart from some exceptional measurements of supersonic flows, these measurements are typically $\leq 40$~km/s.  This was also the range for the cases studied in P03 and G05 where geometrical effects were taken into account.  While such effects may cause flow velocities to be underestimated, many measurements with good alignment exist and these fall within the same range (see also Gontikakis et al.~2006).  Non-thermal velocities, typically 20-40 km/s (Patsourakos \& Klimchuk 2006), tend to be larger than Doppler shifts and are relevant if loops have unresolved sub-structure, with upflows and downflows occurring together in each leg.  Such counter-flows may be typical results of local reconnections between loop strands (Petrie~2006).

The density curves do not differ much from their static counterparts except for cases where $v$ nearly attains the value $\sqrt{g_sZ_0}$, which $\approx 99$~km/s for a tube with this expansion factor and a temperature of 1~MK.  The models of P03 and G05 fall into this category, with adjustments for temperature - this is how they were able to reproduce the scale heights of observations.  While velocity profiles ranging from about 45-100~km/s seem capable of lengthening scale heights considerably, these cases make up a small minority of the solution class, most of which is indistinguishable from the equivalent static class of solutions.  The subclass capable of extending scale heights significantly involves velocities larger than those typically observed in steady loops.  Therefore, unless there is a reason why many loop plasmas should assume such velocity profiles for long periods of time, steady flows do not seem to be responsible for the scale heights routinely observed in active regions.

\section{Discussion}

We have shown that the cross-sectional profile of a coronal magnetic flux tube determines whether steady flow through the tube will increase or reduce the scale height of stratification in the tube.  If tube cross-sectional areas increase sufficiently with height, the scale height would be enhanced by steady flows, in agreement with many observations of super-hydrostatic scale heights in active region loops.  We find that this is likely to occur near active regions where the typical fall-off magnetic field strength with height is estimated to be sufficiently great.  On the other hand, these scale-height extensions are not expected to be significant in the majority of cases since the flow regimes capable of such extensions are rare and involve velocities larger than those typically observed.

Several unresolved questions remain.  Observations repeatedly show that loop emission pattern widths vary little with arclength, in which case steady flows would decrease scale heights.  Near-constant flux tube cross-sections are inconsistent with field strengths decreasing with height, as is well known.  This leads us to question whether these emission patterns really represent flux tubes.

One tentative explanation for emission patterns of constant width is that these patterns delineate loci of magnetic field diffusions which, locally to very narrow regions, heat successive narrow loop strands and accelerate them from these regions during the field reconnections (Petrie 2006).  Dissipations of spontaneous current sheets are expected to occur along flux tube boundaries (Parker~1994), and may well trace out magnetic field trajectories.  If diffusion region sizes, accelerations due to field diffusions and plasma cooling times are uncorrelated with height in the corona, diffusion processes should produce emission patterns of approximately constant width.

Various obstacles stand in the way of determining whether these emission patterns really do represent flux tubes and, if so, of measuring the typical cross-sectional area variation with length of these flux tubes.  It is very difficult to identify entire isolated loops in a single image.  This is partly because length scales across loops are comparable to image pixel sizes, and partly because the observations are taken from a single vantage point.  The second restriction will hopefully be relaxed by the forthcoming STEREO mission.  There is also the question of the distribution of plasma within a loop:  whether loop widths are equally well represented by the emission pattern at all heights.  A flux tube is highly unlikely to have a simple cross-sectional shape in the solar atmosphere, or to maintain a single shape along their entire lengths (Parker~1994, Gudiksen et al.~2005, Lopez Fuentes et al.~2006).  The distortion of a flux tube along its length may greatly influence its appearance from certain angles.  A combined study of active regions using STEREO and a reliable magnetogram source such as GONG or SOLIS will enable us to test for correlations between sharp fall-offs of coronal magnetic field and extended scale heights.  It is hoped that coronal magnetic field measurements (Lin et al.~2004, Tomczyk et al.~2004) will attain sufficiently high spatial resolution for us to check directly whether loop field strengths decrease significantly with height.

For now we can neither reconcile the nearly-constant width of loop emission patterns nor explain extended scale heights in active regions in general.  If active region flux tubes are as expansive as theoretical considerations predict, we have shown that steady flows are at least capable of reproducing the extended heights observed in active regions, but that this only occurs for a small subset of possible flow regimes close to density inversion.  Steady flows are the only steady-state processes known to be capable of extending scale heights.  Either the particular flow regimes capable of this occur often in active regions, or the majority of extended pressure scale heights observed in active regions remain unexplained and there is another process lengthening scale heights that we haven't yet thought of.  The search for such a process remains a challenge for the future and its success would be an important step towards a complete understanding the physics behind the bulk equilibrium of active regions.

\acknowledgements

I thank the referee for constructive comments which led to a substantial improvement of the paper, and Markus Aschwanden for encouragement.  This work was conducted while the author was a participant in the National Aeronautics and Space Administration (NASA) Postdoctoral Program at Goddard Space Flight Center, administered by the National Research Council (NRC) and Oak Ridge Associated Universities (ORAU), and was based at National Solar Observatory, Tucson.

%%%%%%%%%%%%%%%%%%%%%%%%%%%%%%%%%%%%%%%%%%%%%%%%%%%%%%%%%%%%%%%%%%%%%%%%%%%%%%%
%\clearpage

%\appendix
%\section{Figure Captions}

%\noindent
%Figure 1.  

%%%%%%%%%%%%%%%%%%%%%%%%%%%%%%%%%%%%%%%%%%%%%%%%%%%%%%%%%%%%%%%%%%%%%%%%%%%%%%%

%%%%%%%%%%

\clearpage

\clearpage

\end{document}